# Data augmentation in microscopic images for material data mining

Boyuan Ma [a, b, c], Xiaoyan Wei [a, b, c], Chuni Liu [a, b, c], Xiaojuan Ban [a, b, c,*], Haiyou Huang [a, c, d,**], Hao Wang [a, e], Weihua Xue [e, f], Stephen Wu [g,***], Mingfei Gao [a, b, c], Qing Shen [h], Adnan Omer Abuassba [i], Haokai Shen [j, k], Yanjing Su [a, l]

a. Beijing Advanced Innovation Center for Materials Genome Engineering, University of Science and Technology Beijing, Beijing, 100083, China.
b. School of Computer and Communication Engineering, University of Science and Technology Beijing, Beijing, 100083, China.
c. Beijing Key Laboratory of Knowledge Engineering for Materials Science, Beijing, 100083, China.
d. Institute for Advanced Materials and Technology, University of Science and Technology Beijing, Beijing, 100083, China.
e. School of Materials Science and Engineering, University of Science and Technology Beijing, Beijing, 100083, China.
f. School of Materials Science and Technology, Liaoning Technical University, Liaoning, 114051, China.
g. The Institute of Statistical Mathematics, Research Organization of Information and Systems, Tachikawa, Tokyo 190-8562, Japan.
h. National intellectual property administration, Beijing, 100088, China.
i. Arab Open University, Palestine, Ramallah, Palestine.
j. College of Information Science and Engineering, China University of Petroleum, Beijing, China.
k. Key Lab of Petroleum Data Mining, China University of Petroleum, Beijing, China.
l. Corrosion and Protection Center, University of Science and Technology Beijing, Beijing, 100083, China.

* Corresponding author at: School of Computer and Communication Engineering, University of Science and Technology Beijing, Xueyuan Road 30, Haidian District, Beijing 100083, China.
Email addresses: banxj@ustb.edu.cn.
** Corresponding author at: Institute for Advanced Materials and Technology, University of Science and Technology Beijing, Xueyuan Road 30, Haidian District, Beijing 100083, China.
Email addresses: huanghy@mater.ustb.edu.cn.
*** Corresponding author at: Research Organization of Information and Systems, The Institute of Statistical Mathematics, Tachikawa, Tokyo 190-8562, Japan;
Email addresses: stewu@ism.ac.jp.

## Abstract

Recent progress in material data mining has been driven by high-capacity models trained on large datasets. However, collecting experimental data (real data) has been extremely costly since the amount of human effort and expertise required. Here, we develop a novel transfer learning strategy to address small or insufficient data problem. This strategy realizes the fusion of real and simulated data, and the augmentation of training data in data mining procedure. For a specific task of image segmentation, this strategy can generate synthetic images by fusing physical mechanism of simulated images and "image style" of real images. The result shows that the model trained with the acquired synthetic images and 35% of the real images outperforms the model trained on all real images. As the time required to generate synthetic data is almost negligible, this strategy is able to reduce the time cost of real data preparation by roughly 65%.

## Introduction

There has been considerable interest over the last few years in accelerating the process of materials design and discovery [1]. In the past decade, accelerating discovery relied on databases, computation, mathematics and information science has created more and more successful cases in the materials sciences [2-6]. In general, the larger the training dataset is, the more accurate the machine learning model becomes. Especially for deep neural networks, which have shown to have exceptional prediction performance when trained with a large amount of data. As a result, some accelerate methods for data accumulation, such as high-throughput computation and high-throughput experiment, have been developed to build large databases. However, in many materials researches, especially for new materials, we still have to face the dilemma of lacking high-quality data due to a time-consuming or technically difficult process of collecting experimental data (real data) and a low accuracy of computational data.

Material microstructure data is an important type of material data to build the intrinsic relationship of composition, structure, process and properties, which is fundamental to material design. Therefore, the quantitative analysis of microstructures is essential for the control of the properties and performances of metals or alloys [7-8]. One of the most important steps in this process is microscopic image processing using computational algorithms and tools [9-10]. For example, image segmentation [11], which outputs the pixel-wise label of original image, is commonly used to extract significant information in microscopic images at the field of material structure characterization [12]. Take polycrystalline iron for example, as shown in Figure 1(a) and (b), which is a basic and typical case in practical material research, the objective of image processing algorithm is to detect grain boundary from raw microscopic image in order to obtain accurate microstructure information, such as geometric and topological characteristics.

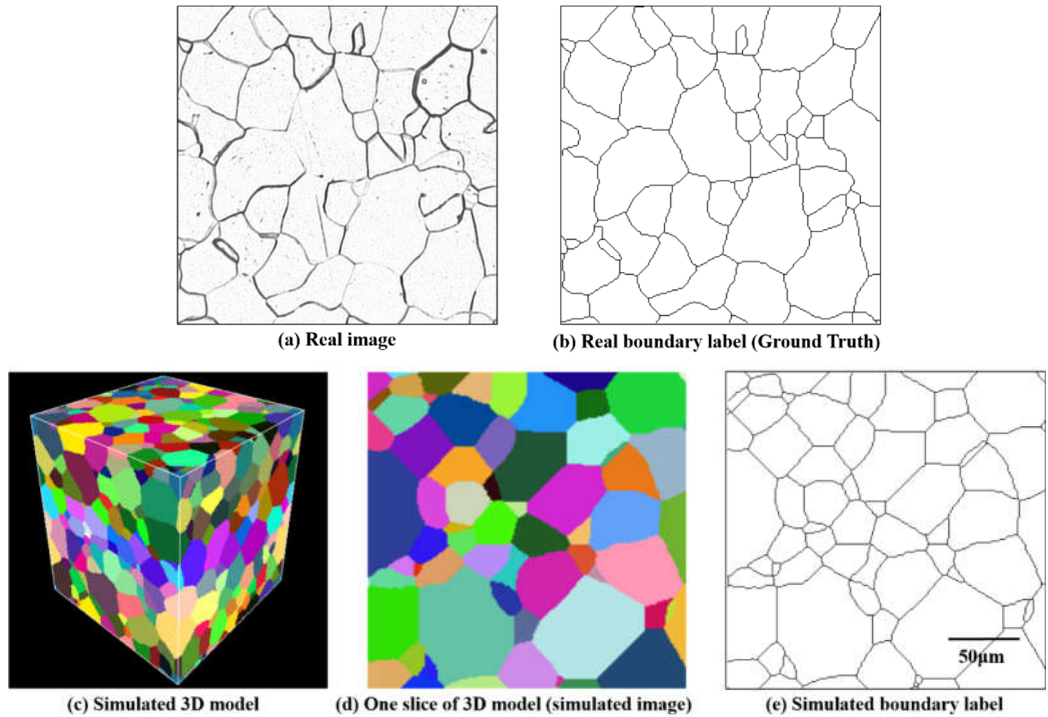

Figure 1. Microscopic images of polycrystalline iron. (a) Raw real experimental image obtained by optical microscope. (b) Grain boundary detection result of (a) conducted by human, it was used as ground truth in our investigation. (c) A 3D simulated model. (d) One slice of image from the 3D model, the so-called simulated image. (e) Boundary detection label of (d).

Recent progress in material microscopic image segmentation [13-15] has been driven by high-capacity models trained on large datasets. Unfortunately, the generalization performance of these models is hindered by the lack of large dataset due to time-consuming labeling of material microscopic images. Creating large datasets with pixel-wise semantic labels is known to be very challenging due to the amount of human effort and expertise required to trace accurate object boundaries [16]. Therefore, such large datasets with pixel-wise label for image segmentation are often hard to obtain.

In this work, we develop a novel transfer learning strategy to address small or insufficient data problem in material data mining. This strategy realizes the fusion of real data and simulated/calculated data based on transfer learning theory, and the augmentation of training data in data mining procedure, so that classification or regression model with better performance can be established based on only a small set of real data.

We explore the use of 3D simulated model to create large-scale pixel-accurate data for training image segmentation systems. For example, as shown in Figure 1(c), a Monte Carlo Potts model can represent polycrystalline microstructure of materials, which turns out to possess similar geometrically and topologically microstructure characteristics of grains compared to the real grain structure [17-18]. It is easy to get large slice (image) data with pixel-level label from simulated 3D model using computational methods, see Figure 1 (d) and (e). However, acquired contents in simulated image data are too perfect, i.e., unrealistic, due to some theoretical approximations and simplifications in the modeling process, which causes a challenge to simply apply simulated image data to real microscopic image processing system. In order to fill in the missing information in simulated image data, we use image-to-image conversion technique to transfer simulated image data into synthetic image which incorporates the information from real images. This image-to-image conversion can be simply described as a model that takes in input of one simulated image, and output a realistic one after processing [19-20]. We are more interested in the fact that the output and input no longer look the same, but the underlying global structure remains unchanged, which is called image style transfer. And, it's shown that conditional Generative Adversarial Networks (condition GANs) do a good job on the task of image style transfer [21-25].

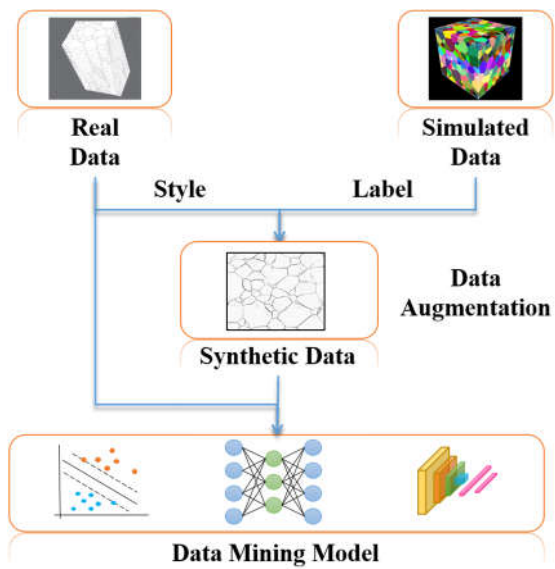

Figure 2. Flowchart of the proposed data augmentation strategy

In general, our presented strategy can create synthetic image data by fusing physical mechanism (pixel-level label) of simulated image data and "image style" of real image data, as shown in Figure 2. The acquired synthetic image is more realistic than simulated image and can be used as a source task in transfer learning. On the quantitative analysis experiment of polycrystalline real image datasets, we show that using the acquired synthetic images increases the performance of image segmentation. In addition, model trained with the acquired synthetic image data and 35% of the real image data outperforms the model trained on all real image data. As the time required to generate synthetic data is almost negligible when compared with the time required to generate real data, our strategy is able to reduce the time cost of real data preparation by roughly 65%. And, we believe that this strategy can be easily applied to other, or even outside materials data mining tasks.

# Results

## Data sets

- **Real image data**

The real image dataset includes a total of 136 serial section optical images of polycrystalline iron with resolution of 2800 × 1600 pixels, of which the ground truth has 2 semantic classes (grain and grain boundary) and labeled by material scientists. It is randomly split into 100 training and 36 test images. The original images were preprocessed into small images (patches) with the size of 400 × 400 pixels, considering the computer capacity and speed. Finally, the training set consists of 2800 patches with size 400 × 400, randomly cut from the original 100 training images, while the test set consists of 1008 patches with size 400 × 400 cut from the other 36 original images directly.

- **Simulated image data**

We establish a large-size 3D simulated model of the polycrystalline materials by Monte Carlo-Potts model. Then we obtain the normal section images from three dimensions, which we called simulated image data, as shown in Figure 1(d). We also calculated the boundary label map, as shown in Figure 1(e). A total of 28800 simulated images were acquired, which are one order of magnitude more than the existing real microscopic image dataset. During the simulation process, we can ensure that the geometric and topological information of the simulated image is statistically consistent with the real image. However, the simulated image data is simple and perfect, i.e., containing only grain boundary information without any "defect" or "noise". For the real image data, the range of pixel values obtained by optical microscope is [0, 255], and the specific pixel value is affected by grain appearance, the light intensity of microscope and noise introduced by sample preparation. For the simulated image data, the range of pixel values is [0, N] with N denotes the number of grains in 3D simulated model, which is controlled by grain-growth simulated model. Therefore, simulated image data can't be directly used in machine learning based algorithm, because there is difference in the nature of two image data.

- **Synthetic image data**

We train our image style transfer model using real image data and simulated boundary label. And we apply our model to transform all the simulated images into synthetic images. As shown in Figure 3, there are 4 columns of images: from left to right is real image, simulated image, simulated boundary label and synthetic image. The synthetic image has both label information and similar "image style" compared with the real image. It can be used as data augmentation for the real image in data mining or machine learning tasks.

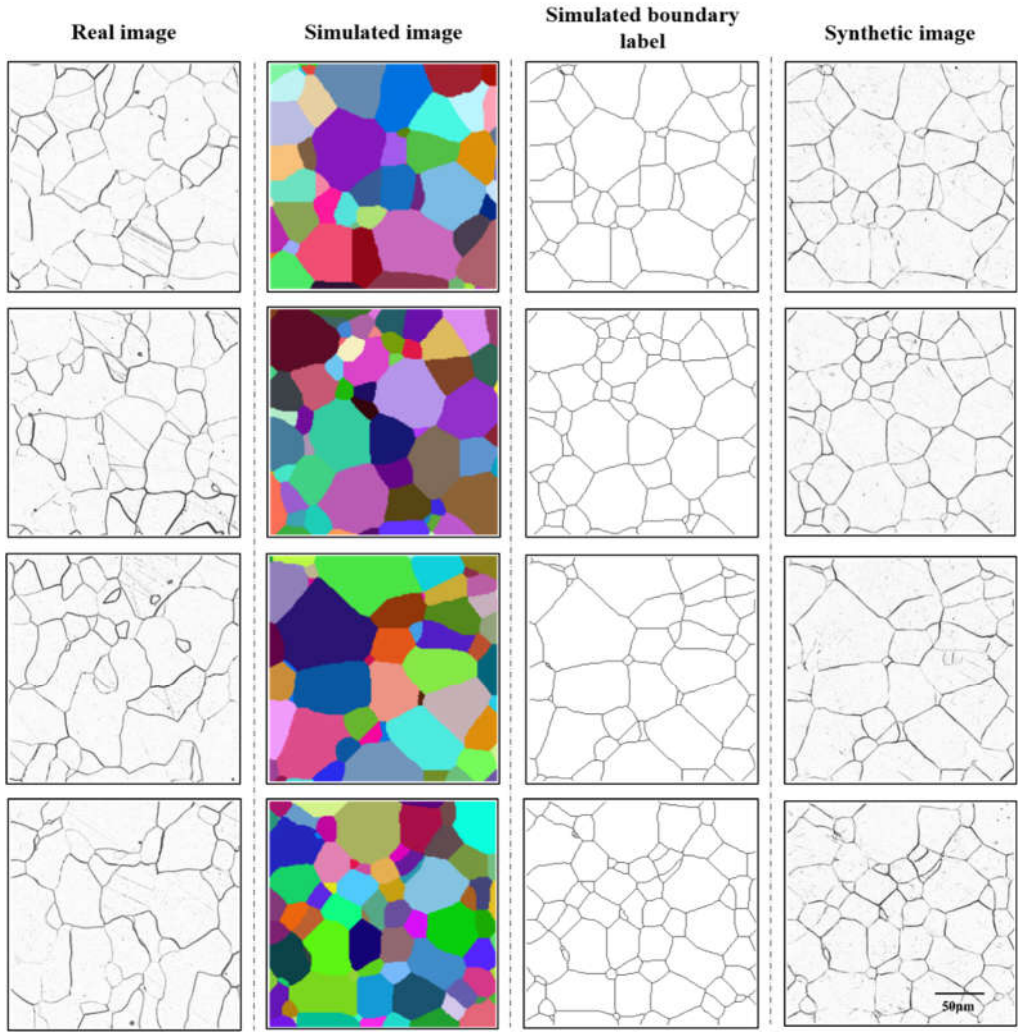

Figure 3. The demonstration of different datasets. From left to right are real image, simulated image, simulated boundary label and synthetic image, respectively

We compare the time consumptions of the three ways of image production, i.e. experiment, simulation and image style transfer in Table 1. Due to complex experimental procedures, including sample preparation, polishing, etching, and photographing, the real image data costs the longest time, about 1200s per image. It should be noted that here we do not consider the increase in time cost caused by failed experiments, that is, the actual experimental process is likely to require more time. While, the preparation of simulated data includes design and establishment of simulated model. By virtue of high-speed computer system, the simulated data costs only one percent of the experimental time, about 12s per image. And the image style transfer takes only 0.1s during the inference of conditional generative adversarial network. Therefore, the time cost of setting up a synthetic image dataset almost depends on the time it takes to obtain experimental images.

Table 1 Manufacture time of three image datasets

| Datasets | Real image data | Simulated image data | Synthetic image data |
|---|---|---|---|
| Operation time (s) per image with the size of 400 × 400 pixels | 1200.00 | 12.00 | 0.10 |

# Evaluation models and metrics

We use U-net [26], an encoder-decoder network, to carry out microscopic image segmentation. In encoder-decoder network, the input goes through a series of convolution-pooling-normalization group layers until the bottleneck layer, where the underlying information is shared with the output. U-net joins the layer-skip connection to transfer the features extracted from the down-sampling layers directly to the upper sampling layers. It makes the pixel location of the network more accurate.

We explore the use of these synthetic images as the data augmentation for the real images. During the training stage, we jointly trained the model on real and synthetic data using batch gradient descent with mini-batches of 8 images, including 4 real and 4 synthetic images, which is same with the work [16]. It takes 28K iterations to converge with a learning rate of $1 \times 10^{-4}$. Although the training pattern is important for network training, this paper don't discuss this because it is beyond our topic. All models are trained with same pattern to ensure fairness. We use two metrics to evaluate our algorithm: Mean Average Precision (MAP) [27-28] and Adjusted Rand Index (ARI) [29-31].

Mean Average Precision (MAP) is a classical measure in image segmentation and object detection task. In this paper, we evaluate it at different intersection over union (IoU) thresholds. The IoU of a proposed set of object pixels and a set of true object pixels is calculated as:

$$IoU(A, B) = \frac{A \cap B}{A \cup B} \quad (1)$$

The metric sweeps over a range of IoU thresholds, at each point calculating an average precision value. The threshold values range from 0.5 to 0.95 with a step size of 0.05: (0.5, 0.55, 0.6, 0.65, 0.7, 0.75, 0.8, 0.85, 0.9, 0.95). In other words, at a threshold of 0.5, a predicted object is considered a "hit" if its intersection over union with a ground truth object is greater than 0.5. Generally, it can be considered that the segment is right when IoU beyond 0.5. The other higher value is aim to ensure the right results.

At each threshold value $t$, a precision value is calculated based on the number of true positives (TP), false negatives (FN), and false positives (FP) resulting from comparing the predicted object to all ground truth objects. And the average precision of a single image is then calculated as the mean of the above precision values at each IoU threshold.

$$Average\ Precision = \frac{1}{|thresholds|} \sum_t \frac{TP(t)}{TP(t) + FP(t) + FN(t)} \quad (2)$$

Finally, the Mean Average Precision (MAP) score returned by the metric is the mean taken over the individual average precisions of each image in the test dataset.

Adjusted Rand index (ARI) is corrected-for-chance version of Rand Index (RI), which is a measure of the similarity between two data clustering's [29-31]. From a mathematical standpoint, ARI or RI is related to the accuracy. Besides, image segmentation can be considered as a clustering task, which split all pixels in images into $n$ partitions or segments.

Given a set $S$ of $n$ elements (pixels), and two groupings or partitions of these elements, namely $X = \{X_1, X_2, \dots, X_r\}$ (a partition of $S$ into $r$ subsets) and $Y = \{Y_1, Y_2, \dots, X_S\}$ (a partition of $S$ into $s$ subsets), the overlap between $X$ and $Y$ can be summarized in a contingency table $[n_{ij}]$ where each entry $n_{ij}$ denotes the number of objects in common between $X_i$ and $Y_j$ : $n_{ij} = |X_i \cap Y_j|$. For image segmentation task, $X$ and $Y$ can be treated as ground truth and predicted result, respectively.

Table 2 Contingency table

|     | $Y_1$ | $Y_2$ | ... | $Y_s$ | Sums |
|-----|-------|-------|-----|-------|------|
| $X_1$ | $n_{11}$ | $n_{12}$ | ... | $n_{1s}$ | $a_1$ |
| $X_2$ | $n_{21}$ | $n_{22}$ | ... | $n_{2s}$ | $a_2$ |
| ... | ... | ... | ... | ... | ... |
| $X_r$ | $n_{r1}$ | $n_{r2}$ | ... | $n_{rs}$ | $a_r$ |
| Sums | $b_1$ | $b_2$ | ... | $b_s$ | |

The ARI is defined as follows:

$$\widetilde{ARI}^{Adjusted\ Rand\ Index} = \frac{\sum_{ij}\binom{n_{ij}}{2} - [\sum_i\binom{a_i}{2}\sum_j\binom{b_j}{2}]/\binom{n}{2}}{\frac{1}{2}[\sum_i\binom{a_i}{2} + \sum_j\binom{b_j}{2}] - [\sum_i\binom{a_i}{2}\sum_j\binom{b_j}{2}]/\binom{n}{2}} \quad (3)$$

Where $n_{ij}$, $a_i$, $b_j$ are values from the contingency table. And $\binom{n}{2}$ is calculated as $n(n-1)/2$.

The MAP is a strict metric because it is mean score on each threshold, so that the score is always smaller than ARI.

For all of them, the higher the metrics, the better the models are. For fair comparison, all the models are evaluated on the same real test set.

Our implementation of this algorithm is derived from the publicly available Python [32], Pytorch framework [33], and OpenCV toolbox [34]. The image style transfer model and U-net's training and testing are performed on a workstation using 4 NVIDIA 1080ti GPU with 44GB memory.

## Image Segmentation by the Proposed Augmentation Method

At first, we explore to use simulated dataset as data augmentation directly. As shown in Table 3, using the real test set, we compare the model's performance which trained on the whole real data, named $Real_{100\%}$ dataset, and that of the whole simulated data, named $Simulated_{100\%}$ dataset. The subscript denotes the percentage of specific dataset used in training set. We find that simply use $Simulated_{100\%}$ dataset achieve poor performance when compared with $Real_{100\%}$ dataset. Besides, the performance will be degraded if we mix these two datasets as training set. We assume that there are two possible reasons. One possibility is that we did not have sufficient simulated data to achieve proper model training. However, additional experiment in supplemental materials shows that models trained with simulated data perform very well on the simulated test set, so that we rule out this possibility. The other reason could be that the simulated data is unrealistic, i.e., containing only grain boundary information without any "defect" that may appear in real data. This problem can be addressed by image style transfer model.

Table 3 The performance of real and simulated data in image segmentation

| Data Sets | MAP | ARI |
|---|---|---|
| $Real_{100\%}$ | **0.5845** | **0.8655** |
| $Simulated_{100\%}$ | 0.1120 | 0.0778 |
| Mixed ($Real_{100\%}$+$Simulated_{100\%}$) | 0.5036 | 0.7522 |

We use image style transfer model to create synthetic image data by fusing pixel level label of simulated image data and "image style" of real image data. We believe that this processing will make simulated data more realistic and could be used as data augmentation for real image data in material data mining.

We evaluate the performance of image style transfer model using segmentation metrics. There are two stages of our approaches, image style transfer and image segmentation, both of them need to be trained. In order to figure out how much real image data is needed to train a promising image style transfer network, we start from small set of real dataset (5%) and increase the amount by 5% each time continuously. For example, we use $Real_{5\%}$ as training set to train an image style transfer model. And then we use the trained image style transfer model to convert all simulated images to the synthetic dataset, called $Synthetic_{100\%}^{5\%}$. The subscript denotes the percentage of specific dataset used in training set and the superscript refers to the ratio of real images to train an image style transfer model.

We show the performance of synthetic data as data augmentation in Experimental results demonstrate that the proposed data augmentation method improves the performance according to quantitative assessment and result visualization.

Table 4. We compare the performance with real dataset and mixed dataset (real and synthetic). We find that the mixed dataset performs better than single real dataset. If we only have 5% of real dataset, the performance will increase about 8% on MAP and 10% on ARI after using synthetic dataset as data augmentation. This suggests that using synthetic dataset as data augmentation would bring significant performance improvement, especially when there's only a small amount of real data.

As shown in Experimental results demonstrate that the proposed data augmentation method improves the performance according to quantitative assessment and result visualization.

Table 4, with the increase of the amount of real data, the performance of model improves for both cases of using only real image training set or mixed training set. When we use 35% of real data, the performance (MAP is 0.5860 and ARI is 0.8749) of mixed training ($Real_{35\%}$+$Synthetic_{100\%}^{35\%}$) outperforms that (MAP is 0.5845 and ARI is 0.8655) of the whole real data ($Real_{100\%}$) on both metrics. It proves that our method can significantly reduce the amount of real data by 65% in image segmentation, which can reduce pressure of getting and labeling real images from experiment.

The visualization of image segmentation is shown in Figure 4. From left to right is real image, real boundary label, the results of model training with $real_{35\%}$ and the results of model training with mixed data ($Real_{35\%}$+ $Synthetic_{100\%}^{35\%}$). The red circles denote the areas that model trained with small real data can't close the grain boundary, but model trained with data augmentation can segment those areas correctly. Experimental results demonstrate that the proposed data augmentation method improves the performance according to quantitative assessment and result visualization.

Table 4 the performance of synthetic data in image segmentation

| Dataset | MAP | ARI | Dataset | MAP | ARI |
| --- | --- | --- | --- | --- | --- |
| $Real_{5\%}$ | 0.4808 | 0.7351 | Mixed($Real_{5\%}$+$Synthetic_{100\%}^{5\%}$) | 0.5599 | 0.8368 |
| $Real_{10\%}$ | 0.5055 | 0.7740 | Mixed($Real_{10\%}$+$Synthetic_{100\%}^{10\%}$) | 0.5721 | 0.8508 |
| $Real_{15\%}$ | 0.5311 | 0.8002 | Mixed($Real_{15\%}$+$Synthetic_{100\%}^{15\%}$) | 0.5742 | 0.8534 |
| $Real_{20\%}$ | 0.5418 | 0.8300 | Mixed($Real_{20\%}$+$Synthetic_{100\%}^{20\%}$) | 0.5823 | 0.8577 |
| $Real_{25\%}$ | 0.5343 | 0.8316 | Mixed($Real_{25\%}$+$Synthetic_{100\%}^{25\%}$) | 0.5792 | 0.8696 |
| $Real_{3\_0\%}$ | 0.5479 | 0.8370 | Mixed($Real_{3\_0\%}$+$Synthetic_{100\%}^{3\_0\%}$) | 0.5863 | 0.8630 |
| $Real_{3\_5\%}$ | 0.5585 | 0.8358 | Mixed($Real_{3\_5\%}$+$Synthetic_{100\%}^{3\_5\%}$) | **0.5860** | **0.8749** |

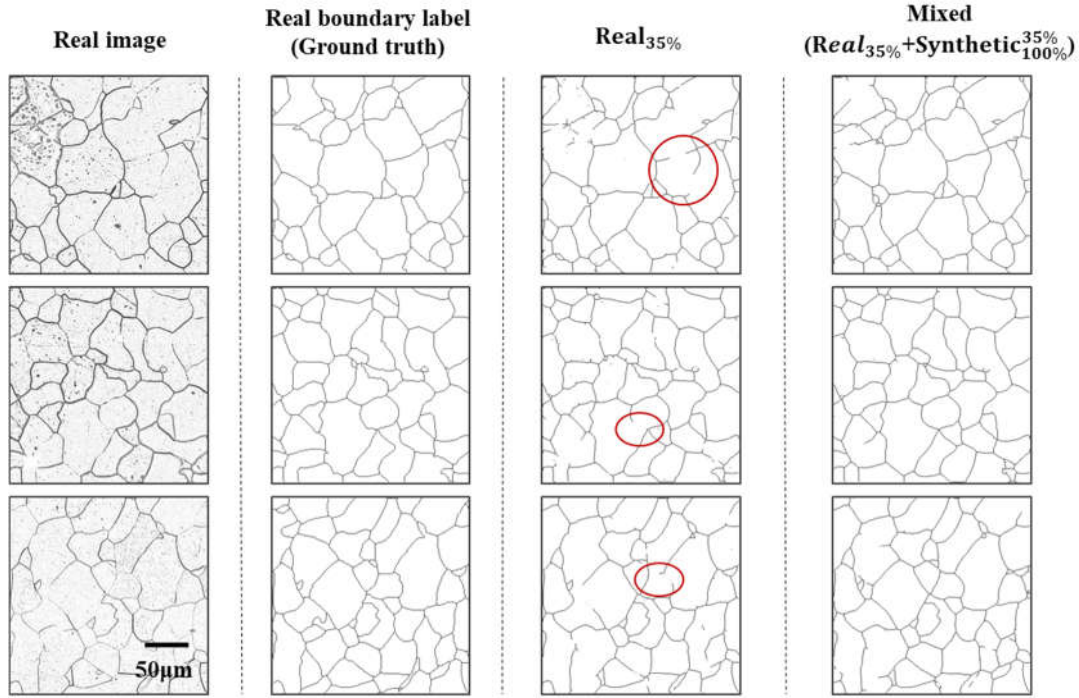

Figure 4 The result of image segmentation with different training set. From left to right is real image, real boundary label, the result of model training with $real_{3\_5\%}$ and the result of model training with mixed data ($real_{3\_5\%}$+ $Synthetic_{100\%}^{3\_5\%}$).

# Discussion

In the past decade, data-driven material modeling has become a popular tool to accelerate material discovery. Generally, Modern machine models (especially deep learning models) have outstanding prediction performance when trained with sufficiently large amount of data. However, for most applications in materials science, there are always a lack of experimental data for a specific task, i.e., the small data dilemma.

In present work, we developed a novel transfer learning strategy to address small or insufficient data problem in material data mining. This strategy realizes the fusion of experimental data (real data) and simulated/calculated data based on transfer learning theory, and the augmentation of training data in data mining procedure, so that classification or regression model with better performance can be established based on only a small set of real data. Then, we applied this strategy to a specific task of image segmentation. First, it transfers "image style" of real experiment image to simulated image in order to generate synthetic image. By fusing physical mechanism of simulated image and "image style" of real image, synthetic image is more realistic than simulated image and can be used as a source task in transfer learning [35]. Second, by supplementing machine learning model with synthetic images, the model captures useful features from simulated data that are transferable to the real data and achieves the promising performance.

Computational simulation is a good way to acquire data efficiently. We assume that Monte

Carlo Potts model could generate simulated data which mimic the pattern of grain growth in ideal condition. In materials science, simulated data partially captures the actual physical mechanism. However, Monte Carlo Potts model could not simulate the perturbation of data in practical experiments. As shown in Table 3, when there is a big difference between the simulated data and the real data, it is not enough to simply mix the simulated data into the real data, which may sometimes lead to a negative effect.

Generative adversarial net, which is the primary architecture behind our image style transfer model, can learn a loss that tries to classify if the output image is real or fake, while simultaneously training a generative model to minimize this loss. We think that, during this processing, the generative model will become more and more powerful that it can learn the small perturbation of data which commonly occurs in actual experiments. Therefore, we could use it generate synthetic data.

By treating synthetic images as data augmentation for machine learning model, it achieves promising performance when trained on a mix of 35% real images and the acquired synthetic images, which is better than the model trained on the whole set of real images. In addition, as the time required to generate synthetic data is almost negligible when compared with the time required to generate real data, our method is able to reduce the time cost of real data preparation by roughly 65%.

This paper has demonstrated the viability of the combination of simulated and real experiment data, suggesting that simulated data (after performing image style transfer to the real data) could be proved useful in data mining or machine learning system. And we believe that the proposed strategy can be easily applied to other, or even outside materials data mining tasks.

Our next step would be to investigate a more efficient transfer learning technique for this segmentation task in order to fully suppress the need of using real experimental data for training.

# Method

## Experimental

A commercial hot-rolled iron plate with purity of >99.9 wt.% was used in this work. The plate was forged into a round bar with a diameter of 30 mm, and then was fully recrystallized by annealing at 1153K for 3 hours. The samples for metallographic characterization were spark cut from the bar. One surface of the samples was mechanically polished for a fixed time and then was taken metallographic photographs with an optical microscope after etched using 4vol% nital solution. The steps of polish-etching-photograph above were repeated to obtain serials section photographs. The average thickness interval is about 1.8μm between two sections.

## Monte Carlo Potts simulated model

Monte Carlo-Potts model [17-18] is used to simulate three-dimensional normal grain growth. We use a 400×400×400 cubic lattice with full periodic boundary conditions to represent the continuum microstructure. A positive integer $S_i$, termed as an index number, is assigned to each site in the lattice sequentially. The index number of a site corresponds to the orientation of the grain that it belongs to. Sites with the same index are considered to be part of the same grain and grain boundary only exists between neighbors with different orientation. The Potts model serves as the grain boundary energy function:

$$E = -J \sum_{i=1}^{N} \sum_{j=1}^{NN} \left( \delta_{S_i S_j} - 1 \right), \quad \delta_{S_i S_j} = \begin{cases} 1, S_i = S_j \\ 0, S_i \neq S_j \end{cases} \quad (4)$$

where E is the boundary energy, N is the system size (the total sites in the system), J is the positive constant which scales the boundary energy, NN is the number of nearest neighbors j of site i, and δ is the Kronecker function with $\delta_{S_i S_j} = 1$ if $S_i = S_j$ and 0 otherwise. Here, NN=26 which means that the interactions between a given site and its 6 first-nearest neighboring sites, 12 second-nearest neighboring sites and 8 third-nearest neighboring sites are considered in the calculation of grain boundary energy. Simulation of grain growth involves reorientation attempt of each site. The reorientation attempt is restricted to the random site which is adjacent to the selected site. Thus the net energy change associated with the reorientation trial can be expressed by:

$$\Delta E_i = E_{i2} - E_{i1} \quad (5)$$

Where $E_{i1}$ and $E_{i2}$ are boundary energy of site i before and after reorientation trial. The possibility of this reorientation is given by:

$$W = \begin{cases} 1, & \Delta E \leq 0 \\ \exp\left(-\frac{\Delta E}{k_B T}\right), & \Delta E > 0 \end{cases} \quad (6)$$

where $k_B$ is the Boltzmann constant and T is the Monte Carlo temperature. The term $k_B T$ defines the thermal fluctuations within the simulated system. Time is scaled by Monte Carlo step (MCS) which corresponds to as many orientation trials as there are lattice sites. The following shows the iterative procedure of Monte Carlo simulation of the normal grain growth:

1) Each site is selected as a target at the same time.
2) One of the sites in the range of the first-nearest and second-nearest neighbors of the target is randomly selected and the orientation of the selected neighboring site is referred to a trial orientation of the target.

The probability of the trial reorientation for target is determined by eq. (3) with the constant $\left(\frac{k_B T}{J}\right) = 0.5$, which is large enough to reduce lattice pinning, but small enough to avoid lattice break-up.

Figure 1 (c) shows the Monte Carlo simulation structure. Figure 5 shows the distribution of grain area and the distribution of the number of grain edge. It can be seen that pure iron structure and Monte Carlo simulation structure has similar features in grain morphology, grain area distribution and edge number distribution.

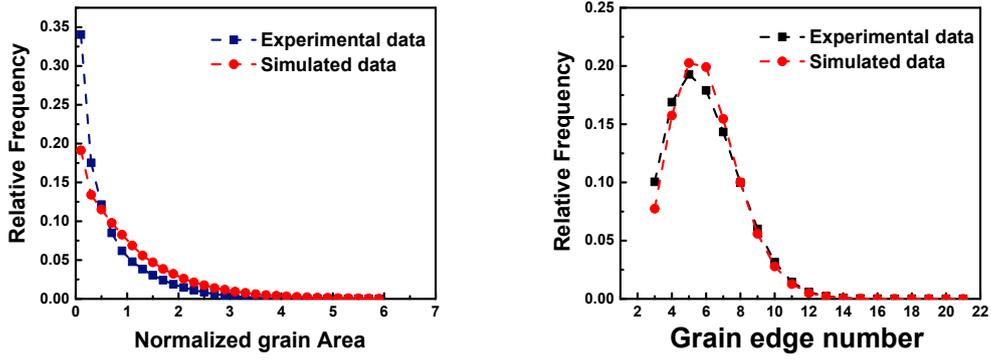

Figure 5. (a) the distribution of grain area (b) the distribution of grain edge

## Image Style Transfer

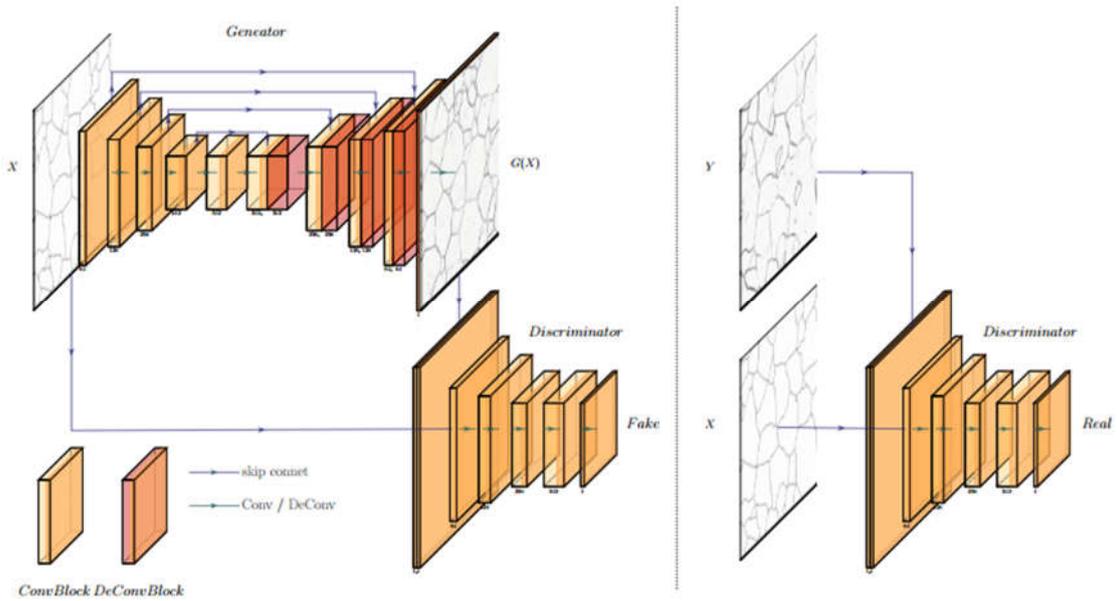

Figure 6. The structure of style transfer model. The primary model behind this model is generative adversarial net. The discriminator D learns to distinguish $G(x)$ from y, the generator G learns to fool the D.

We use image style transfer algorithm to make simulated image data acquiring the "image style" of real image data, as shown in Figure 6, where x is simulated boundary label, $G(x)$ is synthetic result and y is the real image. Specifically, we use conditional GANs to carry out transformation, the so-called pix2pix [21] [36]. Conditional GANs convert image x to image y. The given image x is called the condition, as the input of the generator G. During training, the generator G produces the output $G(x)$. While, the discriminator D distinguish $G(x)$ from y. Both modules are optimized by adversarial training to make "fake" $G(x)$ closer to "true" $y$ More importantly, the output $G(x)$ must retain the underlying structural similarity to the condition x. In other word, this condition can forge the net to retain the label of the simulated image.

The objective function of a conditional GAN can be expressed as:

$$\mathbf{L_{cGAN}(G, D)} = E_{x,y}[log\, D(x,y)] + E_x[\log(1 - D(x, G(x)))] \quad (7)$$

where G tries to minimize this objective, while D tries to maximize it.

$$\mathbf{G^* = arg\, min_G\, max_D\, L_{cGAN}(G, D)} \quad (8)$$

The article [19] explored that generator has to not only fool the discriminator but also be near the ground truth output in an L1 sense, using L1 distance to mix the objective. The L1 loss is described as bellow:

$$L_1(G) = E_{x,y}[\|y - G(x)\|_1] \qquad (9)$$

The final objective is:

$$G^* = \arg\min_G \max_D L_{cGAN}(G, D) + \lambda L_1(G) \qquad (10)$$

In order to realize conversion from simulated images to real microscopic images, our generator uses the encoding-decoding network U-net [26]. And that network structure enables the input and output to be shared on the bottleneck layer, which helps to retain underlying structural similarity. The discriminator calculates the loss of local patches between output and ground truth to represent the consistency of high-level details.

## Data availability

The datasets generated and/or analyzed during the current study are available from the corresponding author on reasonable request. And the third-party image style transfer codes that we used are available online [36].

# Acknowledge

The authors acknowledge the financial support from the National Key Research and Development Program of China (No. 2016YFB0700500), and the National Science Foundation of China (No. 51574027, No. 61572075, No. 6170203, No. 61873299), and Key Research Plan of Hainan Province (No. ZDYF2018139), and Fundamental Research Funds for the University of Science and Technology Beijing (No. FRF-BD-19-012A).

# Author contributions

B.M. conceived the idea, designed the experiment and wrote the paper; X.W participated in paper writing and conducted experiment with C.L.; H.H. participated in experimental designing and discussion with S.W.; H.W. and W.X. prepared the data for the experiments. M.G. prepared the computational environment of experiment. Q.S., A.O.A. and H.S. were involved in the analyses of data. X.B and Y.S coordinated research project and provided financial support.

# Competing interests

B.M., X.W., X.B., H.H., H.W. and W.X. declare the following competing interests that one patent has been registered (201910243002.6).

# Supplemental Materials

In experimental section, we find that the model trained with whole simulated data achieve very poor performance on real test set. We assume that there are two possible reasons. One is that insufficient simulated data might lead to insufficient model training. Therefore, we conduct an experiment using simulated data as training set and testing set. The simulated train set and simulated test set were divided from the whole simulated data according to the proportion of 2.8:1, consistent with the partitioning ratio of real data sets. As shown in Table 1, the MAP and ARI are very high which means that the simulated data is enough to train a promising model. Besides, both values are near to 1, which denotes that the simulated data is too perfect and simple. Thus, simulated data needs to be added image style from real image before it use as data augmentation for material data mining.

Table 1 The image segmentation performance of simulated data in the simulated test set

| Data Sets | MAP | ARI |
| --- | --- | --- |
| Simulated (for training and testing) | 0.9591 | 0.9986 |